\documentclass[fleqn,usenatbib]{mnras}

\usepackage[T1]{fontenc}

\DeclareRobustCommand{\VAN}[3]{#2}
\let\VANthebibliography\thebibliography
\def\thebibliography{\DeclareRobustCommand{\VAN}[3]{##3}\VANthebibliography}

\usepackage{graphicx}	% Including figure files
\usepackage{amsmath}	% Advanced maths commands
\usepackage{amssymb}	% Extra maths symbols
\title[$CO_{\mathrm{2}}$ condensation on rocky planets]{The influence of surface $CO_{\mathrm{2}}$ condensation on the evolution of warm and cold rocky planets orbiting Sun-like stars}

\author[I. Bonati and R. Ramirez]{
I. Bonati,$^{1}$\thanks{E-mail: irene.bonati@elsi.jp (IB)}
R. M. Ramirez,$^{1,2}$
\\
$^{1}$Earth-Life Science Institute, Tokyo Institute of Technology, Tokyo, Japan\\
$^{2}$Space Science Institute, Boulder, Colorado,USA\\
}

\date{Accepted Accepted 2021 March 19. Received 2021 March 17; in original form 2020 April 13.}
\pubyear{2021}
\usepackage{newtxtext,newtxmath}
\begin{document}
\definecolor{greenblue}{RGB}{0,127,127}
\definecolor{greenish}{RGB}{104, 159, 56}
\newcommand{\add}[1]{\textcolor{greenblue}{#1}}
\newcommand{\irene}[1]{\textcolor{greenblue}{\textit{(Irene: #1)}}}
\newcommand{\ramses}[1]{\textcolor{greenish}{\textit{(Ramses: #1)}}}
\newcommand{\todo}[1]{\textcolor{red}{#1}}
\newcommand{\textb}[1]{\textcolor{greenblue}{#1}}
\newcommand{\textg}[1]{\textcolor{greenish}{#1}}
\label{firstpage}
\pagerange{\pageref{firstpage}--\pageref{lastpage}}
\maketitle

% Abstract of the paper
\begin{abstract}
The habitable zone is the region around a star where standing bodies of liquid water can be stable on a planetary surface. Its width is often assumed to be dictated by the efficiency of the carbonate-silicate cycle, which has maintained habitable surface conditions on our planet for billions of years. This cycle may be inhibited by surface condensation of significant amounts of $CO_{\mathrm{2}}$ ice, which is likely to occur on distant planets containing high enough levels of atmospheric $CO_{\mathrm{2}}$. Such a process could permanently trap $CO_{\mathrm{2}}$ ice within the planet, threatening its long-term habitability. Recent work has modeled this scenario for initially cold and icy planetary bodies orbiting the Sun. Here, we use an advanced energy balance model to consider both initially warm and cold rapidly-rotating planets orbiting F - K stars. We show that the range of orbital distances where significant surface $CO_{\mathrm{2}}$ ice condensation occurs is significantly reduced for warm start planets. Star type does not affect this conclusion, although surface $CO_{\mathrm{2}}$ ice condenses over a larger fraction of the habitable zone around hotter stars. The warm start simulations are thus consistent with 1-D model predictions, suggesting that the classical habitable zone limits in those earlier models are still valid. We also find that the cold start simulations exhibit trends that are consistent with those of previous work for the Sun although we now extend the analysis to other star types.    
\end{abstract}


%%%%%%%%%%%%%%%%%%%%%%%%%%%%%%%%%%%%%%%%%%%%%%%%%%

%%%%%%%%%%%%%%%%% BODY OF PAPER %%%%%%%%%%%%%%%%%%

\section{Introduction}

During the course of its evolution, Earth is thought to have exhibited mostly warm surface conditions that were temporarily interrupted by a few sporadic cooling episodes, triggered by processes taking place in its interior, atmosphere, and host star. The resultant climatic changes can manifest as mild oscillations, or even as extreme snowball events capable of engulfing Earth's entire surface in ice \citep{kirschvink1992,Hoffman1342}. Nevertheless, the carbonate-silicate cycle, which maintains the balance of atmospheric $CO_{\mathrm{2}}$ between the atmosphere and the interior, has allowed Earth to escape permanent glaciation and maintain clement surface conditions throughout time \citep{Hoffman1342}.

\begin{figure}
	\includegraphics[width=0.83\columnwidth]{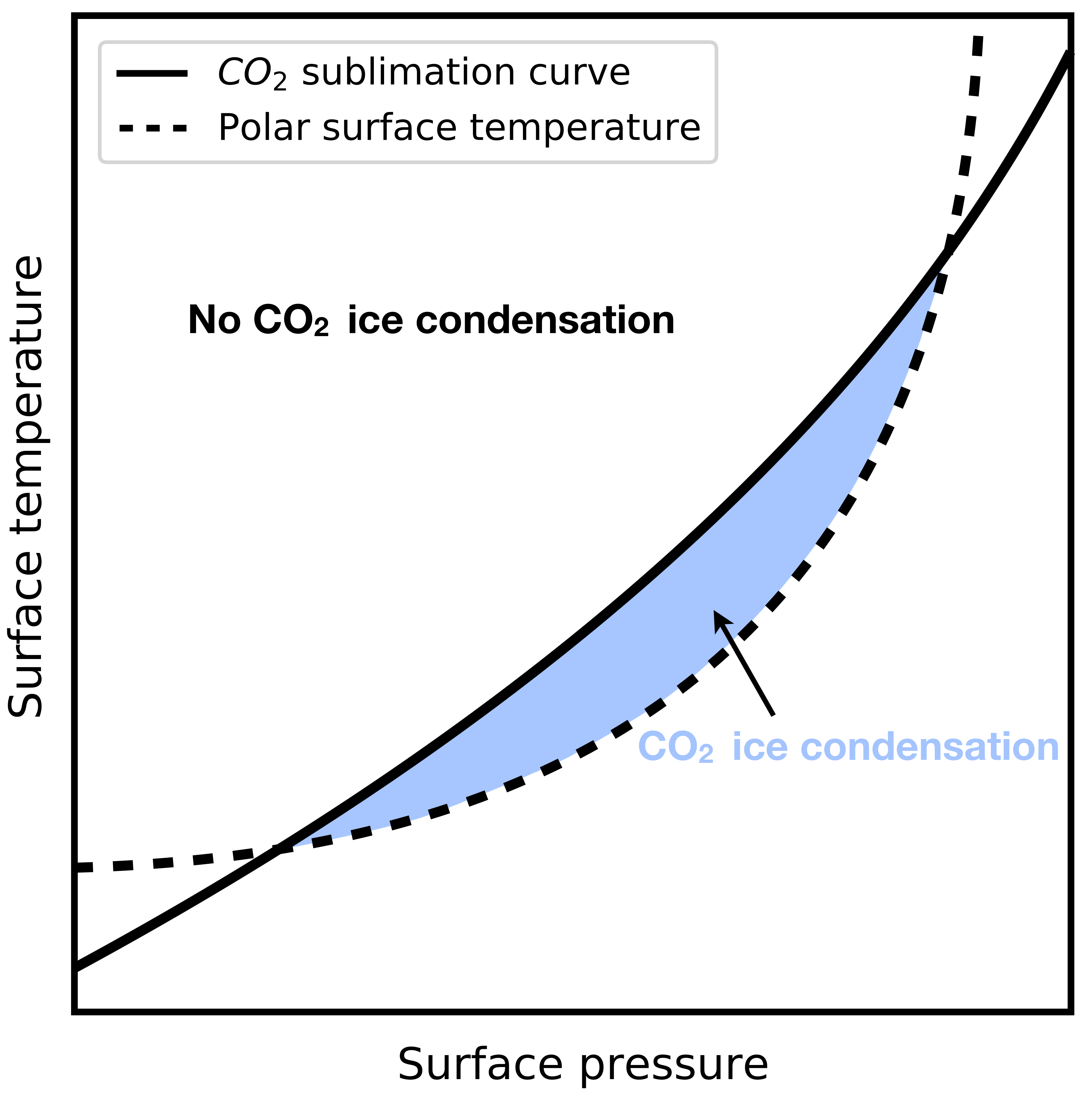}
    \caption{Schematic representation showing the $CO_{\mathrm{2}}$ sublimation and the polar surface temperature curves as a function of surface temperature and pressure. At the poles, the surface temperature is low enough to drop below the $CO_{\mathrm{2}}$ sublimation temperature. The presence of a sufficiently high amount of atmospheric $CO_{\mathrm{2}}$ can, in turn, drive precipitation of $CO_{\mathrm{2}}$ ice onto the planetary surface. 
    Such a phenomenon is less likely to happen for a small atmospheric $CO_{\mathrm{2}}$ inventory or when the greenhouse effect of $CO_{\mathrm{2}}$ dominates at sufficiently high surface pressures. Adapted from Figure 1 of \citet{Soto2015}.}
    \label{fig:Sketch}
\end{figure}

However, large amounts of atmospheric $CO_{\mathrm{2}}$ on a cold enough planet, as is the case for some depictions of early Mars \citep{Kasting1991,wordsworth2013}, can have a detrimental effect on its habitability. A number of recent studies \citep{Kasting1991,Phumbert2005,Phumbert2011, forget2013,Soto2015, Turbet2017,kadoya_outer_2019} have shown that once $CO_{\mathrm{2}}$ partial pressures exceed the saturation level, and surface temperatures are low enough, significant condensation of $CO_{\rm 2}$ ice can occur at the poles (Figure~\ref{fig:Sketch}). The relatively high Rayleigh scattering of $CO_{\mathrm{2}}$, along with the resultant increase in the planetary albedo, work to enhance the ice-albedo feedback \citep{Kasting1991} and keep the planet cold. Moreover, the high density of $CO_{\mathrm{2}}$ ice compared to water ice can lead to surface condensation and subsurface sequestration of the former, removing some $CO_{\mathrm{2}}$ from the atmosphere forever. This could result in a permanently glaciated state and prevent planets from achieving warm states. In that case,  the influence of the carbonate-silicate cycle in sustaining habitability is greatly lessened, if not gone altogether \citep{Turbet2017}.  

While the carbonate-silicate cycle has stabilized surface temperatures over time on Earth, this might not be the case for bodies orbiting other stars and/or located at different orbital distances, and having different atmospheric $CO_{\mathrm{2}}$ inventories. In particular, $CO_{\mathrm{2}}$ pressures at large orbital distances are high enough for surface condensation and Rayleigh scattering to counter the greenhouse effect, directly impacting the width of the classical $CO_{\mathrm{2}}$-$H_{\mathrm{2}}O$ habitable zone \citep{kasting1993,KumarKopparapu2013, Ramirez2018}. It is therefore important to assess the role of $CO_{\mathrm{2}}$ in determining a given planet's fate. \citet{Turbet2017} have recently addressed this issue by using a 3-D Global Climate Model (GCM) to simulate the evolution of terrestrial planets across the habitable zone, and concluded that bodies initially assumed to be fully frozen remained that way, exhibiting permanent surface $CO_{\mathrm{2}}$ ice condensation at orbital distances as small as \textasciitilde $1.27$~AU from the Sun for an atmospheric $CO_{\mathrm{2}}$ pressure of $1$~bar. In comparison, 1-D calculations suggest that the habitable zone extends to \textasciitilde $1.47$~AU at that $CO_{\mathrm{2}}$ pressure level \citep{kasting1993,KumarKopparapu2013, Ramirez2018}. 
All of this suggests that $CO_{\mathrm{2}}$ condensation in more advanced climate modeling simulations may have an even more detrimental effect on planetary habitability than previously calculated, possibly decreasing the size of the habitable zone as compared to 1-D modeling simulations.

\citet{Turbet2017} had simulated planets with initially fully-glaciated surface conditions (cold start worlds) orbiting the Sun. In contrast, habitable-zone planets that start their evolution warm (warm start worlds) may exhibit a different response to $CO_{\mathrm{2}}$ condensation. Here, we employ an advanced energy balance model (EBM) to investigate the parameter space of surface $CO_{\mathrm{2}}$ condensation for cold and warm rotating planets with different atmospheric $CO_{\mathrm{2}}$ pressures orbiting Sun-like (F - K) stars. We compare the cold and warm start results and discuss implications for the width of the habitable zone.

\section{Methods}
\label{sec:Methods}
\subsection{Governing equations}
We make use of an advanced non-grey energy balance model (EBM), similar to that described in \citet{RamirezLevi2018} and \citet{ramirez2020}, which is itself based on  other similar models (see also \citet{North1979},  \citet{North1981}, \citet{Williams1997}, and \citet{vladilo2013}) to determine the presence of $CO_{\mathrm{2}}$ ice condensation and its influence on the fate of bodies orbiting Sun-like stars. We study the evolution of Earth-like planets assuming cold ($T_{\mathrm{surf}}$=230 K) or warm ($T_{\mathrm{surf}}$=280 K) starts. The EBM is coupled to a 1-D radiative-convective climate model that provides the radiative transfer calculations. The reader is redirected to \citet{ramirez2020} for a more detailed explanation of the model, but we reiterate some of the key details here.

The present model follows the radiative energy balance principle \citep[e.g.,][]{Williams1997}, according to which planets in thermal equilibrium radiate as much energy to space as they receive from their host star. The atmospheric-ocean energy balance is expressed as \citep[e.g.,][]{james1982,Williams1997,batalha2016}: 

\begin{equation}
\label{ebmeq}
    C_{\mathrm{p}} \frac{\partial T(x, t)}{\partial t}-\frac{\partial}{\partial x} D\left(1-x^{2}\right) \frac{\partial T(x, t)}{\partial x}+I-L\left (\frac{dM_{\mathrm{col,CO_{\mathrm{2}}}}}{dt}\right )=S(1-A),
\end{equation}{}
where $C_{\mathrm{p}}$ is the effective heat capacity, $T$ is the zonally averaged surface temperature, $x$ is the sine of the latitude, $t$ is time, $D$ is the heat diffusion coefficient (i.e., the latitudinal transport of heat), $I$ is the outgoing infrared flux to space, $S$ is the absorbed fraction of incident solar flux, $L$ is the latent heat flux per unit mass of $CO_{\mathrm{2}}$ (5.9 \ $\cdot$ 10\textsuperscript{5}J/kg; \citet{forget1998}), $M_{\mathrm{col,CO_{\mathrm{2}}}}$ is the column mass of $CO_{\mathrm{2}}$ which sublimates from or condenses onto the planetary surface, and $A$ is the albedo at the top of the atmosphere. Equation \ref{ebmeq} is solved for every time step using a second-order finite difference scheme.

The modeled planets are Earth-sized and are subdivided into 36 five degree wide latitudinal belts with land and ocean coverage similar to present-day Earth (i.e., $70 \%$ oceans and $30 \%$ land). Flat topography is assumed for simplicity. Planets are in circular orbits and the length of day is 24 hours. As per \citet{Turbet2017}, we assume that volcanic outgassing rates at a given stellar flux are always high enough to support stable climates. Thus, our worlds do not undergo limit cycles, which lead to hypothetical oscillations between warm and cold climates for planets with low volcanic outgassing rates \citep{haqq2016limit,paradise2017,kadoya_outer_2019}. We note that the existence of limit cycles is controversial and is based on various weathering and surface albedo assumptions that may not be applicable in many circumstances \citep{ramirez2017mars,graham-a}.

Our calculations assume that atmospheres are fully-saturated and consist of $1$~bar $N_{\mathrm{2}}$ for a range of atmospheric $CO_{\mathrm{2}}$ pressures. The thermal diffusion coefficient $D$ is calculated using the scaling relation (with the subscript "$0$" referring to present Earth values):

\begin{equation}
\label{hdiff}
    \left(\frac{D}{D_{\mathrm{0}}}\right)=\left(\frac{p}{p_{\mathrm{0}}}\right)\left(\frac{C_{\mathrm{p}}}{C_{\mathrm{p}_{0}}}\right)\left(\frac{m_{\mathrm{0}}}{m}\right)^{2}\left(\frac{\Omega_{0}}{\Omega}\right)^{2}\left(\frac{T_{\mathrm{surf,ave}}}{T_{\mathrm{0}}}\right)^{12},
\end{equation}
where $D_{\mathrm{0}}=0.58$~Wm\textsuperscript{-2}K\textsuperscript{-1}, $C_{\mathrm{p0}}=10^{3}$~g\textsuperscript{-1}kgK\textsuperscript{-1}, $p$ is the atmospheric pressure ($p_{\mathrm{0}}=1$~bar), $m$ is the atmospheric mass ($m_{\mathrm{0}}=28$~kg), $\Omega$ is the planetary rotation rate ($\Omega_{\mathrm{0}}=7.27 \cdot 10^{\mathrm{-5}}$~rads\textsuperscript{-1}), and $T_{\mathrm{surf,ave}}$ is the annual average surface temperature ($T_{\mathrm{0}}=288$~K). A crude temperature dependence is added to this equation to simulate the importance of latent heat release and transport at high temperatures \citep{Caballero2005a,Rose2017}. The exponent $12$ on the right side of Equation~\ref{hdiff} gives the right latitudinal temperature structure in comparing our results to those of Figure 1 from \citet{Turbet2017}, which displays an average temperature $T_{\mathrm{surf,ave}} \sim 225$~K. Without this temperature dependence, equator-pole temperature gradients are underestimated and heat transport is overestimated at cold temperatures.

The model is able to distinguish between land, ocean, ice, and clouds. As the atmosphere warms near and above the freezing point, water clouds form, with the latitudinal cloud coverage ($c$) dictated by:
\begin{equation}
\label{cloud_coverage}
    c = \min \left ( 0.72 \log \left( \frac{F_{\mathrm{C}}}{F_{\mathrm{E}}} + 1 \right ),1 \right ).
\end{equation}
Here, $F_\mathrm{C}$ is the convective heat flux, $F_\mathrm{E}$ is the convective heat flux for the Earth at $288$~K ($\sim 90$~W/m$^2$). This equation is similar to that used in the Community Atmosphere Model (CAM) GCM \citep{xu1991, yang2014} and yields an Earth-like cloud coverage $c\sim 50 \%$ at a mean surface temperature of $288$~K. For simplicity, cloud decks do not overlap\add[]{,} and an averaged cloud fraction is computed at each latitude. This is an improvement over 1-D models that place such clouds at the surface \citep{kasting1993, KumarKopparapu2013}. 

The $CO_{\rm{2}}$ clouds modeled here are non-absorbing, and thus radiatively inactive. This is consistent with recent simulations suggesting that their greenhouse effect may be very small in these dense $CO_{\rm{2}}$ atmospheres \citep{kitzmann2016}. Even so, our $CO_{\rm{2}}$ clouds still impact the planetary albedo, affecting planetary surface temperatures and $CO_{\rm{2}}$ precipitation. Following GCM predictions \citep{forget2013}, a cloud coverage $c=50 \%$ is assumed for $CO_{\rm{2}}$. Such clouds form once temperatures along the adiabat are cold enough for $CO_{\rm{2}}$ to condense. Following \citep{Williams1997} we assume that all $CO_{\rm{2}}$ clouds form in that layer. 

For each time step, spanning about $2$ hours of a planet's evolution, the new average surface temperature is updated for every latitude belt following Equation~\ref{ebmeq}, along with the resulting $H_{\mathrm{2}}O$ and $CO_{\mathrm{2}}$ inventories in the atmosphere and surface. The model simulates the entire year, including seasons, using an explicit forward marching numerical scheme to achieve convergence. Such convergence is reached when the average annual surface temperature does not change by more than \textasciitilde$0.1$~K and both $CO_{\mathrm{2}}$ condensation and sublimation rates are balanced. 

\begin{figure}
	\includegraphics[width=\columnwidth]{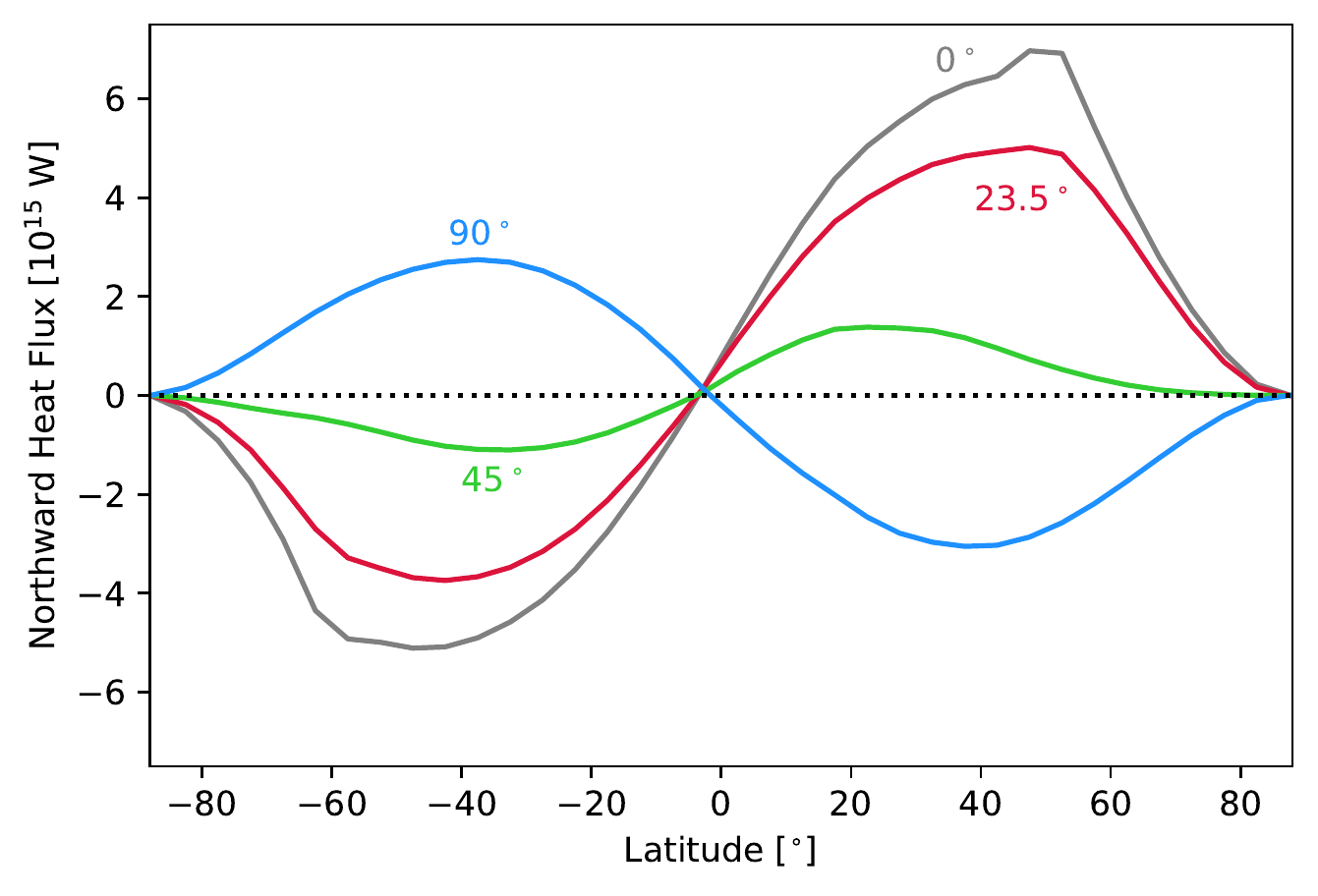}
    \caption{Meridional heat flow for an Earth-like planet  orbiting a Sun-like star at $1$~AU, having an atmospheric pressure of $CO_{\mathrm{2}}$ of $3.3\cdot10^{-4}$~bar and obliquities $0^\circ$, $23.5^\circ$, $45^\circ$, and $90^\circ$. At low obliquities, heat is transported from the equator towards the poles, while transport in the opposite direction is favored at high enough obliquities.}
    \label{fig:flow}
\end{figure}

To validate our model, we compute the northward heat flux for a planet with $330$~ppm $CO_{\mathrm{2}}$ (i.e., similar to present Earth) that orbits the Sun at $1$~AU at different obliquities by using the following expression:

\begin{equation}
    \mathcal{F}_{\mathrm{\lambda}}=2 \pi R \cos \lambda F_{\mathrm{\lambda}}=2 \pi R^{2} D \cos ^{2} \lambda \frac{\partial T}{\partial x},
\end{equation}

where $\mathcal{F}_{\mathrm{\lambda}}$ is the latitudinal energy transport per unit length, $R$ is the planetary radius, $\lambda$ is the latitude (between $-90^\circ$ and $90^\circ$, and $\frac{\partial T}{\partial x}$ is the temperature gradient between different latitude belts.
The obtained flux, shown in Figure~\ref{fig:flow}, matches well with similar models used in past studies \citep[e.g.,][]{WilliamsPollard}.

At the start of the final orbit (after convergence is achieved), we track how much surface $CO_{\mathrm{2}}$  condenses into ice, melts, or sublimates at a given latitude and time step. This approach provides a consistent comparison between warm and cold start scenarios. The phase curve of $CO_{\mathrm{2}}$ employed in this model uses established data parameterized from previous 1-D radiative-convective climate models \citep{Kasting1991} and matches very well with modern data \citep{fray_sublimation_2009}. $CO_{\rm 2}$ ice forms at cold enough surface temperatures and when the saturation pressure is reached.  The albedo of surface $CO_{\rm 2}$ ice (dry ice) in our model is 0.6 \citep{warren_spectral_1990}, which is the same value used in \citet{Turbet2017}.

\subsection{Initial conditions and parameter space}
We model the evolution of Earth-sized bodies, orbiting  F0, K5, and solar (G2) stars. We vary the initial $CO_{\mathrm{2}}$ atmospheric pressure, surface temperature (cold and warm start), planetary obliquity, and semi-major axis. The initial conditions are summarized in Table~\ref{tab:ic}. If a planet starts out cold, there are initially no water clouds in the atmosphere. The global surface temperature is set to $T_{\mathrm{surf}}=230$~K, and the land snow fraction is equal to $1$ (i.e., the continents are fully covered in ice). This was done to approximate the initial conditions of \citet{Turbet2017}.

On the other hand, if a planet starts out warm, the surface temperature is set to $T_{\mathrm{surf}}=280$ K, with a land snow fraction of $0.5$. The fractional cloud cover from water vapor is $c_{\mathrm{H_{\mathrm{2}}O}}=0.26$ (resulting from Equation~\ref{ebmeq}). 
There are no $CO_{\mathrm{2}}$ clouds in the initial step for either case (i.e., $c_{\mathrm{CO_{\mathrm{2}}}}=0$). These starting conditions are meant to approximately simulate the modeling conditions of \citet{Turbet2017}. However, after the initial step, both cloud coverage and surface temperatures are calculated self-consistently. 

We explore different parameters as shown in Table~\ref{tab:parameters}. We do not simulate planets located at larger semi-major axis distances than those showed in Table~\ref{tab:parameters} because modeled surface temperatures at these $CO_{\mathrm{2}}$ pressures are below what is possible for our radiative transfer model ($T_{\mathrm{surf}}=150$~K). Nevertheless, the sampled parameter space is more than sufficient to obtain the overall trends.

\begin{table}
	\centering
	\caption{Initial conditions for planets starting out cold and warm.}
	\label{tab:ic}
	\begin{tabular}{ccc} 
		\hline
		Parameter & \textbf{Cold start} & \textbf{Warm start} \\
		\hline
		Surface temperature $T_{\mathrm{surf}}$ [K] & 230 & 280 \\
		$H_{\mathrm{2}}O$ cloud fraction $c_{\mathrm{H_{\mathrm{2}}O}}$ & 0 & 0.26  \\
		$CO_{\mathrm{2}}$ cloud fraction $c_{\mathrm{CO_{\mathrm{2}}}}$ & 0 & 0 \\
		Land snow fraction & 1 & 0.5 \\
		\hline
	\end{tabular}
\end{table}

\begin{table}
	\centering
	\caption{Parameter space investigated in the simulations.}
	\label{tab:parameters}
	\begin{tabular}{ccccc}
		\hline
		&\multicolumn{1}{c}{Parameter}&\multicolumn{3}{c}{Values}\\
		\hline
		&\multicolumn{1}{c}{$CO_{\mathrm{2}}$ partial pressure range [bar]}&\multicolumn{3}{c}{0.01-3.0}\\
		&\multicolumn{1}{c}{Obliquity [$^{\circ}$] }&\multicolumn{3}{c}{0, 23.5}\\
		&\multicolumn{1}{c}{}&\multicolumn{1}{c}{\underline{F0 star}}&\multicolumn{1}{c}{\underline{Sun}}&\multicolumn{1}{c}{\underline{K5 star}}\\
		&\multicolumn{1}{c}{Stellar temperature [K]}&\multicolumn{1}{c}{7200}&\multicolumn{1}{c}{5800}&\multicolumn{1}{c}{4400}\\
		&\multicolumn{1}{c}{Stellar mass [$M_{\odot}$]}&\multicolumn{1}{c}{1.5}&\multicolumn{1}{c}{1.0}&\multicolumn{1}{c}{0.6}\\
		&\multicolumn{1}{c}{Stellar luminosity [$L_{\odot}$]}&\multicolumn{1}{c}{4.3}&\multicolumn{1}{c}{1.0}&\multicolumn{1}{c}{0.15}\\
		&\multicolumn{1}{c}{Semi-major axis range [AU]}&\multicolumn{1}{c}{2.0-3.0}&\multicolumn{1}{c}{1.0-1.5}&\multicolumn{1}{c}{0.4-0.6}\\
		\hline
	\end{tabular}
\end{table}

\section{Results}

\subsection{Steady-state climate regimes}

\begin{figure*}
	\includegraphics[width=\textwidth]{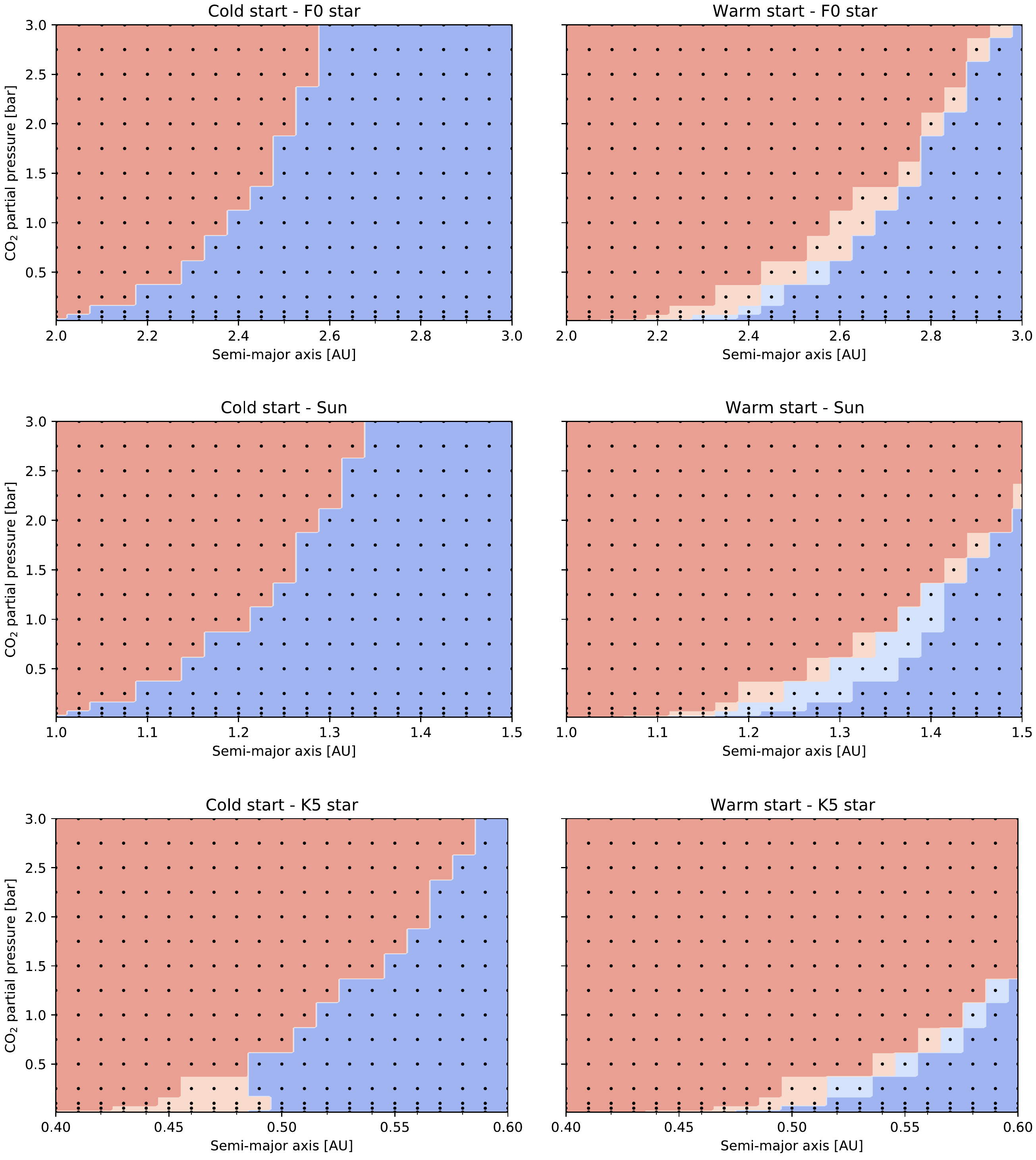}
    \caption{Steady-state solutions reached by warm start (ice-free body) and cold start (fully-glaciated body) planets orbiting F0, Sun-like, and K5-type stars as a function of orbital distance (in AU) and initial atmospheric $CO_{\mathrm{2}}$ partial pressure. The modeled planets have an Earth-like continental fraction and an obliquity of $23.5^{\circ}$. The dark and light red regions comprise bodies that are ice-free and partially (or fully) covered in $H_{\mathrm{2}}O$ ice, respectively. The light and dark blue regions represent partially (or fully) ice-covered and snowball planets, respectively, exhibiting both $H_{\mathrm{2}}O$ and $CO_{\mathrm{2}}$ ice on their surfaces. Note: The horizontal scale is different for each host star. Our model predicts similar trends at $0^{\circ}$ obliquity (not shown).} 
    \label{fig:ss_all}
\end{figure*}

We allow the $CO_{\mathrm{2}}$ partial pressure to evolve until $CO_{\mathrm{2}}$ sublimation and condensation rates are balanced. Figure~\ref{fig:ss_all} shows the steady-state climate regimes obtained for initially cold and warm Earth-sized bodies having an obliquity of $23.5^{\circ}$ and orbiting different types of stars (solar twin-G2, as well as F0 and K5 stars), as a function of their orbital distance and initial atmospheric $CO_{\mathrm{2}}$ pressure. We distinguish between four final scenarios (Figure~\ref{fig:ss_all}). 

As expected, surface $CO_{\mathrm{2}}$ ice condensation starts occurring at smaller orbital distances for cold start planets (left panels in Figure~\ref{fig:ss_all}) as compared to bodies that start out warm (right panels in Figure~\ref{fig:ss_all}).
For the solar case, the $1$~bar $CO_{\mathrm{2}}$ cold start scenario exhibits surface $CO_{\mathrm{2}}$ ice condensation starting from orbital distances as small as $\sim1.22$~AU, which compares favorably to the $\sim 1.27$~AU value found by \citet{Turbet2017}. The equivalent distances for the F0 and K5 stars are $\sim2.4$~AU and $\sim0.52$~AU, respectively.
In contrast, surface $CO_{\mathrm{2}}$ ice condensation for warm starts does not occur until $\sim1.37$~AU for the $1$~bar $CO_{\mathrm{2}}$ solar case, $0.15$~AU farther out than for the cold start scenario. At this same pressure level, $CO_{\mathrm{2}}$ surface ice starts forming at $\sim2.7$~AU and $\sim0.58$~AU for the F0 and K5 stars, respectively. This is about $0.3$~AU and $0.06$~AU farther out than for the equivalent cold start cases.
For comparison, 1-D radiative-convective climate modeling simulations predict that planets with a $1$~bar $CO_{\mathrm{2}}$ atmosphere orbiting F0, solar, and K5 stars can display habitable surface conditions at distances as far as $\sim 2.75$, $1.47$, and $0.58$~AU, respectively \citep{kasting1993, KumarKopparapu2013,Ramirez2018}, assuming the luminosity values given in Table~\ref{tab:parameters}. These values are only slightly farther out than the most distant extent of the red regions at the $1$~bar level (Figure~\ref{fig:ss_all}). Although nearly the same size, the warm regions in the warm starts still span slightly smaller areas (generally) than predicted in 1-D radiative-convective climate modeling simulations \citep{kasting1993, KumarKopparapu2013,Ramirez2018}. This is because of the EBM's increased ice-albedo feedback.

The distances beyond which surface $CO_{\mathrm{2}}$ ice condensation starts forming are pressure-dependent, as a result of the greenhouse effect of $CO_{\mathrm{2}}$. At higher pressures, surface $CO_{\mathrm{2}}$ ice condensation in both warm and cold start cases occurs farther away from the host star, whereas the opposite occurs at lower pressures. Moreover, we find that the surface $CO_{\mathrm{2}}$ ice condensation encompasses a larger region for the hotter stars (alternatively, the dark and light red areas are larger for cooler stars). This is because near-infrared absorption is lower and Rayleigh scattering is higher for planetary atmospheres orbiting hotter stars, cooling the planet and causing $CO_{\mathrm{2}}$ surface ice condensation to occur closer to the star.

A key observation from our cold starts is that a planet cannot escape full glaciation for orbital distances exceeding $\sim 2.6$, $\sim1.33$, and $\sim0.59$~AU for the F0, solar, and K5 star cold start scenarios (Figure~\ref{fig:ss_all}). In these cases, the planet remains glaciated with surface $CO_{\mathrm{2}}$ ice condensation regardless of the atmospheric $CO_{\mathrm{2}}$ content. This difference is attributed to the weaker heat transport and higher surface albedo associated with such cold starts. Although \citet{Turbet2017} does not necessarily obtain $CO_{\mathrm{2}}$ surface ice in all of these cases, in agreement with our model, they predict snowball states. 

\subsection{Latitudinal variation of surface temperature and amount of surface $CO_{\rm 2}$ ice}
\label{sec:surf_temp}

\begin{figure*}
	\includegraphics[width=0.93\textwidth]{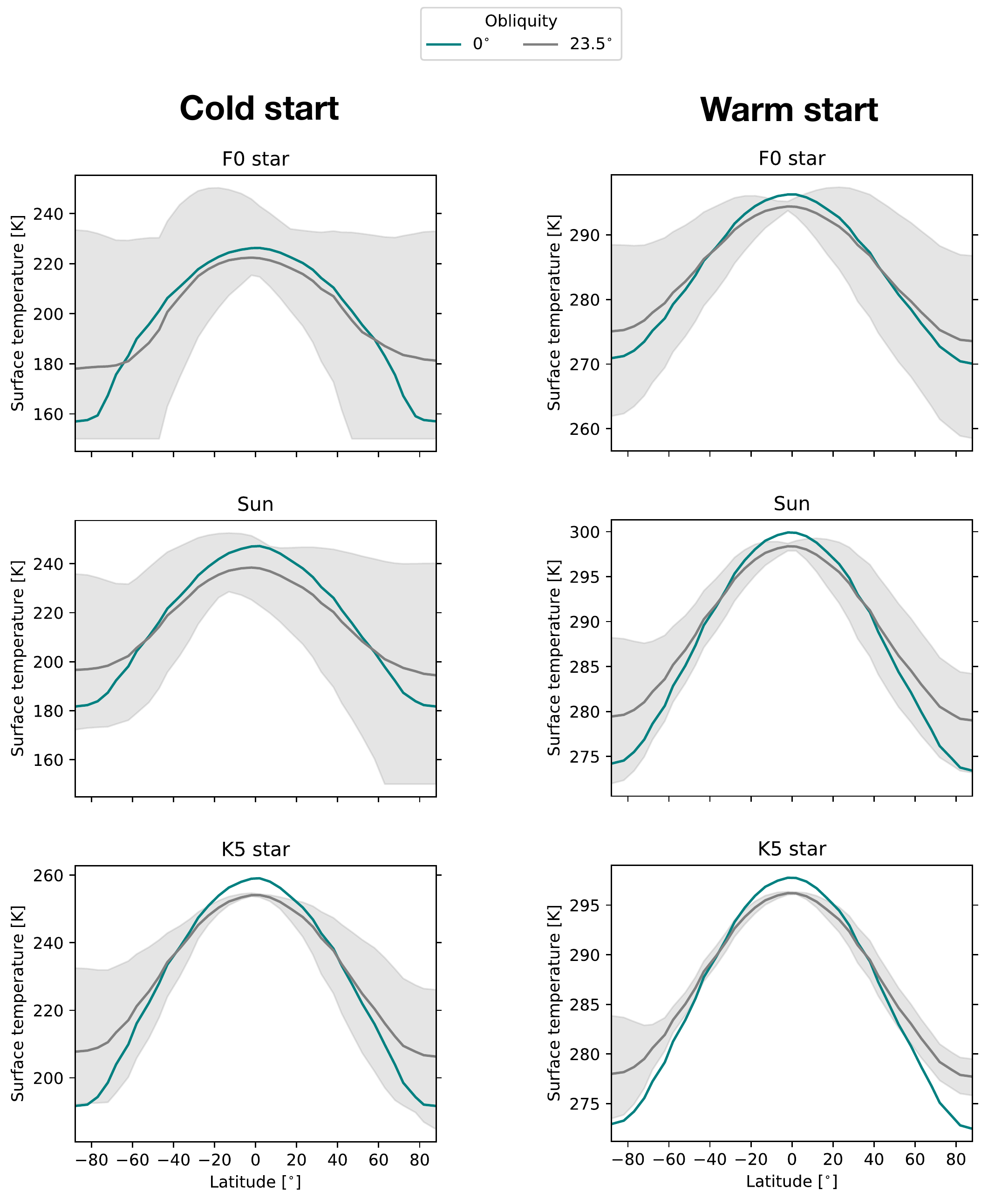}
    \caption{Latitudinally averaged annual surface temperature profiles for cold start (initial $T_{\mathrm{surf}}=230$~K; left column) and warm start (initial $T_{\mathrm{surf}}=280$~K; right column) planets with an initial atmospheric $CO_{\mathrm{2}}$ pressure of $1$~bar for both $0^{\circ}$ (dark green) and $23.5^{\circ}$ (grey) obliquity cases orbiting F0, K5, and Sun-like stars at $2.5$~AU, $0.53$~AU, and $1.25$~AU, respectively. Solid lines represent mean annual surface temperatures, and the shaded regions comprise the amplitude of variation between the minimum and maximum annual surface temperatures for $23.5^{\circ}$ obliquity. In contrast, $0^{\circ}$ obliquity planets experience no seasonal variability. Warm start planets (right column) always have an atmospheric $CO_{\rm 2}$ pressure of $1$~bar. For fully glaciated planets (left column) the annual average amounts of atmospheric $CO_{\rm 2}$ are $6.91\cdot 10^{-2}$~bar ($0^{\circ}$) and $4.22\cdot 10^{-2}$~bar ($23.5^{\circ}$) for bodies orbiting F0 stars, $3.96\cdot 10^{-1}$~bar ($0^{\circ}$) and $2.49\cdot 10^{-1}$~bar ($23.5^{\circ}$) for planets orbiting Sun-like stars, and $8.35\cdot 10^{-1}$~bar ($0^{\circ}$) and $8.94\cdot 10^{-1}$~bar ($23.5^{\circ}$) for planets orbiting K5 stars. The remaining $CO_{\rm 2}$ condenses on the planetary surface as ice. Seasonal ice may persist at the end of the model year, which is the close of southern winter. The model does not compute latitudinal temperatures below $T_{\rm surf}=150$ K, although calculated surface temperatures are mostly above that threshold. The few cases that do not are seasonal minimum temperatures that do not significantly impact average temperatures or computed trends.
    }
    \label{fig:comp_a0}
\end{figure*}

The surface temperatures during the last orbit are illustrated in Figure~\ref{fig:comp_a0} as a function of latitude, for both initially cold and warm planets having an initial atmospheric $CO_{\mathrm{2}}$ pressure of $1$~bar and obliquities of $23.5^{\circ}$ and $0^{\circ}$. These planets are located at intermediate orbital distances from the host star, close to the transition between ice-free and fully ice-covered bodies (see Figure \ref{fig:ss_all}). At such semi-major axes (and intermediate atmospheric $CO_{\rm 2}$ pressures), the initial state of a planet (warm or cold) is a strong predictor of the final state of the planet. For instance, warm start planets tend to end up warm as well. Likewise, a cold start planet at these distances converges to a frozen solution. In contrast, at smaller orbital distances, the intense starlight produces warm planets regardless of the starting state. At larger semi-major axes, the reduction in starlight, combined with more intense surface $CO_{\rm 2}$ condensation, favors cold solutions (Figure \ref{fig:ss_all}). 

Low obliquity bodies receive more direct insulation at the equator, which causes large temperature differences between the latter and the poles, as well as generally lower polar surface temperatures. This also produces the tendency for equatorial surface temperatures to be higher at $0^{\circ}$ than at $23.5^{\circ}$ obliquities, especially at higher $CO_{\mathrm{2}}$ pressures. For intermediate atmospheric $CO_{\rm 2}$ pressures, this could imply slightly more surface $CO_{\mathrm{2}}$ ice at higher ($23.5^{\circ}$) obliquity. For example, the fraction of atmospheric $CO_{\rm 2}$ that precipitates onto the planetary surface of planets having an initial atmospheric $CO_{\rm 2}$ pressure of $1$~bar is $\sim 93.1 \%$ (0$^{\circ}$ obliquity) and $\sim 95.8\%$ (23.5$^{\circ}$ obliquity) for bodies orbiting F0 stars, $\sim 60.4 \%$ (0$^{\circ}$ obliquity) and $\sim 75.1\%$ (23.5$^{\circ}$ obliquity) for planets surrounding the Sun. \citet{Soto2015} also found a similar trend at these intermediate $CO_{\rm 2}$ pressures. We attribute this to the decreased heat transport in the winter hemisphere, which favors surface $CO_{\rm 2}$ ice formation at the winter poles for these moderately dense atmospheres. For example, meridional heat exchange to the poles is greater at low obliquity than it is to the northern pole during southern summer at high obliquity. This is because the northern pole is farther away from the sub-stellar point (during southern summer) in the high obliquity scenario, producing colder seasonal minimum temperatures. Likewise, seasonal temperatures at the southern pole are at a maximum. The opposite thought experiment holds true for northern summer. Nevertheless, this trend breaks in our model for bodies orbiting K5 stars, with greater condensation at low obliquity: $\sim 16.5 \%$ (0$^{\circ}$ obliquity) and $\sim 10.6\%$ (23.5$^{\circ}$ obliquity) (Figure \ref{fig:comp_a0}). Moreover, like the results by \citet{Soto2015} for the Sun, the trends reverse and our model predicts slightly more surface ice formation at low  $CO_{\rm 2}$ pressures (as low as 0.01 bar; not shown) for the low obliquity solar and F0 cases. Again, the K5 case exhibits the opposite trend, with more surface ice production at high obliquity in that scenario. This discrepancy may be related to differences in the amount of near-infrared emission among the star types. The addition of topography would further complicate these trends, creating cold traps that favor $CO_{\rm 2}$ condensation beyond what is predicted in the flat topography worlds considered here \citep{forget2013, wordsworth2013}. 

 We also find that the amount of condensed surface $CO_{\rm 2}$ at a given equivalent distance (e.g., similar flux level) is highest for planets orbiting F0 stars and lowest for worlds around K5 stars (Figure \ref{fig:comp_a0}). This is because planets orbiting cooler stars exhibit less Rayleigh scattering and enhanced near-infrared absorption, both of which enhance warming \citep{kasting1993}. This suggests that planets orbiting cooler stars may be more resistant against atmospheric collapse.

\subsection{Fraction of condensed $CO_{\rm 2}$ ice}
\begin{figure*}
	\includegraphics[width=\textwidth]{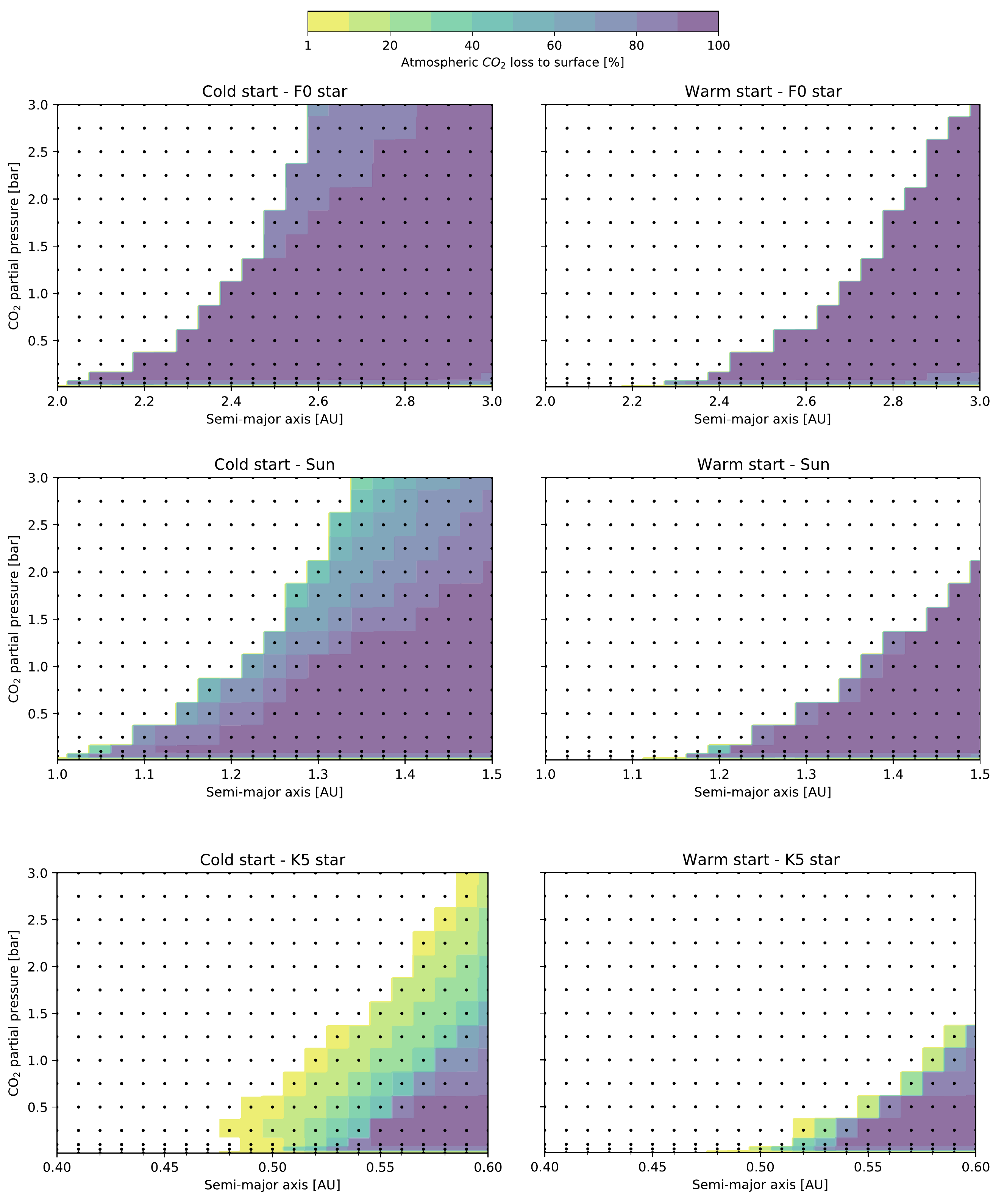}
    \caption{Fraction of atmospheric $CO_{\rm 2}$ condensed on planetary surface as a function of orbital distance (in AU) and initial atmospheric $CO_{\rm 2}$ pressure for cold start (left column) and warm start (right column) planets orbiting F0 (upper row), Sun-like (middle row) and K5 (lower row) stars. The obliquity is $23.5^{\circ}$. White regions denote parameter combinations for which no $CO_{\rm 2}$ ice condensation is observed (see also Figure~\ref{fig:ss_all}).}
    \label{fig:CO2_iceloss}
\end{figure*}

As shown by the Mars simulations of \citet{Soto2015}, the fraction of atmospheric $CO_{\mathrm{2}}$ that condenses onto a planetary surface depends heavily on the initial atmospheric $CO_{\mathrm{2}}$ pressure. The fraction of atmospheric $CO_{\rm 2}$ lost to the surface as a function of orbital distance and starting atmospheric $CO_{\rm 2}$ pressure is shown in Figure~\ref{fig:CO2_iceloss}. The obtained trends reflect the competition between cooling ($CO_{\rm 2}$ ice condensation, clouds, and Rayleigh scattering) and warming (greenhouse effect, stellar insolation) processes. 

We obtain fractions of condensed atmospheric $CO_{\rm 2}$ ranging between $\sim 1\%$ and more than $90\%$. The largest amounts of condensed $CO_{\rm 2}$ are obtained for cold planets located at large orbital distances (e.g., reduced insolation). We note that at the lowest atmospheric $CO_{\mathrm{2}}$ pressures (i.e., smaller than $0.05$~bar) saturation is not reached as easily, and atmospheric $CO_{\mathrm{2}}$ precipitates in smaller amounts (between $1$ and $10\%$, Figure~\ref{fig:CO2_iceloss}). Similarly, at high enough atmospheric $CO_{\mathrm{2}}$ pressures, 
the greenhouse effect dominates, countering surface ice condensation, especially at equatorial latitudes. In particular, it is the enhanced warming from $CO_{\rm 2}$-$CO_{\rm 2}$ collision-induced absorption at higher pressures that offsets such atmospheric losses \citep{wordsworth2010, ramirez2014}. However, while high initial $CO_{\rm 2}$ pressures can decrease the magnitude of condensation at small orbital distances, atmospheric collapse occurs throughout most or even the whole $CO_{\rm 2}$ pressure range explored here for planets on larger orbits. 

Even though the orbital and pressure range at which $CO_{\rm 2}$ condensation takes place is larger for planets that begin their evolution cold (see also Figure~\ref{fig:ss_all}), we find that the transition towards collapse ($>10$~\% atmospheric $CO_{\rm 2}$ condensation) generally occurs more abruptly for warm start planets (Figure~\ref{fig:CO2_iceloss}). This is related to the orbital distance at which $CO_{\mathrm{2}}$ first starts to condense. For the cold start cases, the onset of $CO_{\mathrm{2}}$ condensation occurs closer to the host star, where more insolation is available to counter this process, slowing down the transition to collapse. In contrast, atmospheric $CO_{\mathrm{2}}$ condensation initiates at large orbital distances for warm start planets, where there is less stellar energy to counter condensation.

\section{Discussion}

\subsection{Implications for planetary habitability on early Mars, Earth, and exoplanets}

In agreement with \citet{Turbet2017}, our model finds that $CO_{\mathrm{2}}$ surface condensation can be a detriment to the habitability of cold start planets (Figure~\ref{fig:ss_all}). As the dense surface $CO_{\mathrm{2}}$ ice becomes sequestered within the subsurface, the atmospheric $CO_{\mathrm{2}}$ pressure decreases, which promotes even more ice formation, possibly leading to atmospheric collapse \citep{Turbet2017}. Unlike Earth, where volcanism ended snowball episodes \citep{Hoffman1342}, it might be more difficult for distant cold planets to avoid permanently glaciated conditions. This is because once $CO_{\mathrm{2}}$ pressures exceed saturation, $CO_{\mathrm{2}}$ is increasingly removed from the atmosphere before temperatures ever become warm enough. An exception to this could be volcanism rich in $H_{\mathrm{2}}$ or $CH_{\mathrm{4}}$, which could produce sufficient $CO_{\mathrm{2}}$-$H_{\mathrm{2}}$ or $CO_{\mathrm{2}}$-$CH_{\mathrm{4}}$ collision-induced absorption at high enough concentrations (percent level) and $CO_{\mathrm{2}}$ pressures \citep{ramirez2014,wordsworth_transient_2017,ramirez2017,ramirezkalt2018,turbet_far_2019}. Nevertheless, cold start planets that are close enough to their stars receive enough energy to circumvent the above problems (Figure~\ref{fig:ss_all}). 

The above provides some interesting implications. A number of studies argue that Mars (located at $1.52$~AU) was a cold planet that had undergone numerous transient warming episodes over geologic timescales, possibly aided by supplementary volcanic gases or other mechanisms in a predominantly $CO_{\mathrm{2}}$ atmosphere \citep{wordsworth2013,batalha2016,wordsworth_transient_2017,kite_methane_2020,hayworth_warming_2020}. However, multiple sporadic warming episodes would be very difficult to achieve in practice because the excess $CO_{\mathrm{2}}$ will be removed from the atmosphere once a warm period ends. This will not only enhance the ice-albedo feedback, raising the planetary albedo and triggering atmospheric collapse, but the atmospheric $CO_{\mathrm{2}}$ would be gradually removed from the atmospheric-surface system forever as it sinks below the less dense $H_{\mathrm{2}}O$ ice. Some estimates of the early water inventory suggest that early Mars could have had a global equivalent water layer that was at least a couple of hundred meters deep \citep{villanueva2015,ramirezetal2020}. This would have been a sufficiently large reservoir to absorb much of the condensing surface $CO_{\mathrm{2}}$, decreasing the likelihood of subsequent transient warming episodes \citep{Turbet2017}. 

In the case of limit cycles for instance, models often assume dirty low albedo (0.35) surface $CO_{\mathrm{2}}$ ice \citep{batalha2016, kadoya_outer_2019,hayworth_warming_2020}. This helps increase the absorbed stellar flux, which favors deglaciation and the occurrence of limit cycles. However, observations suggest that the mean $CO_{\mathrm{2}}$ ice albedo on present Mars is much higher \citep{forget2013}. Therefore, assuming such a low albedo on a global scale may be unrealistic. Also, limit cycle models almost always assume a linear relationship between weathering rate and dissolution of [H+] in groundwater \citep{batalha2016,kadoya_outer_2019,hayworth_warming_2020}. However, experiments have shown that this relationship is much weaker for real silicate rocks \citep{asolekar1991}, greatly reducing the occurrence of limit cycles \citep{ramirez2017mars}. Thus, it is not clear if such global episodic warm episodes had ever occurred on early Mars, at least during the climate optimum of maximum fluvial incision. If they did, they may have been very few in number.  

In contrast to our cold start cases, our warm start simulations suggest that the condensation of surface $CO_{\mathrm{2}}$ ice affects a much smaller range of orbital distances. Irrespective of the star type, the range of semi-major axes and $CO_{\mathrm{2}}$ pressures for which planets exhibit habitable surface conditions is similar to that predicted by 1-D calculations of the habitable zone \citep{kasting1993, Ramirez2018}. Warm start cases that are distant enough to manifest significant surface $CO_{\mathrm{2}}$ condensation suffer the same habitability problems discussed above for cold start planets. These results also have implications for our own planet. The Earth, by definition, is located close enough to the star where surface $CO_{\mathrm{2}}$ ice condensation is impossible under normal circumstances (Figure~\ref{fig:ss_all}), and it would have been warm so long as the greenhouse effect, including from $CO_{\mathrm{2}}$, was potent enough. Nevertheless, the Hadean Earth likely had a $CO_{\mathrm{2}}$-rich atmosphere \citep{kasting2014}, suggesting that it was almost certainly a warm start planet, possibly facilitating an early emergence of life.

\subsection{Comparison with previous studies}

Although we obtain similar trends, our cold start results differ in certain respects from those of \citet{Turbet2017}. In particular, there are discrepancies in the spatial coverage of the red (ice-free or ice-covered planets with no $CO_{\mathrm{2}}$ ice) and blue (cold and icy planets displaying surface $CO_{\mathrm{2}}$ ice) colored regions in Figure~\ref{fig:ss_all} between the two studies, which are likely to be mostly attributed to model differences (although there are slight differences in definition of the regions as well). In particular, EBMs such as this one are quite sensitive to the ice-albedo feedback, and are thus more prone to abrupt transitions between cold and warm climate states \citep{RamirezLevi2018}, as shown in Figure~\ref{fig:ss_all}. Our model also employs an atmospheric-ocean heat transport parameterization that likely produces different model behavior during the transition between icy and warm climates than the static ocean assumption of \citet{Turbet2017}. It is unclear how GCMs with fully-coupled atmospheric oceans may compare with the results presented here. This would be an interesting consideration for future studies. 

We note that at low $CO_{\rm 2}$ partial pressures surface $CO_{\rm 2}$ condensation is more easily reached than what is qualitatively shown in Figure~\ref{fig:Sketch}. However, when planets are located far away enough from the host star and/or surface temperatures are low enough (such as for cold starts), $CO_{\rm 2}$ condensation can easily take place during the evolution of the planet, once saturation pressure is reached. Some of the difference may be due to $H_{\rm 2}O$ ice cloud warming that some GCMs predict in dry and cold atmospheres, but our model does not. An alternative reason for the differences may be in the spatial and vertical distributions of clouds, which are not accurately predicted by EBMs. Furthermore, the $CO_{\mathrm{2}}$ clouds formed in our simulations impact the planetary albedo, but they are non-absorbing (see Section~\ref{sec:Methods}). In contrast, the $CO_{\mathrm{2}}$ clouds of \citet{Turbet2017} are either radiatively active or inactive across the entire spectrum. Nevertheless, other GCM results are consistent with 1-D results (and, in turn, the EBM results here in the location of the outer edge of the habitable zone; \citet{wolf2018erratum}). Our warm start results agree well with those predicted from 1-D radiative-convective climate models \citep{kasting1993, KumarKopparapu2013, Ramirez2018}.

\begin{figure}
	\includegraphics[width=\columnwidth]{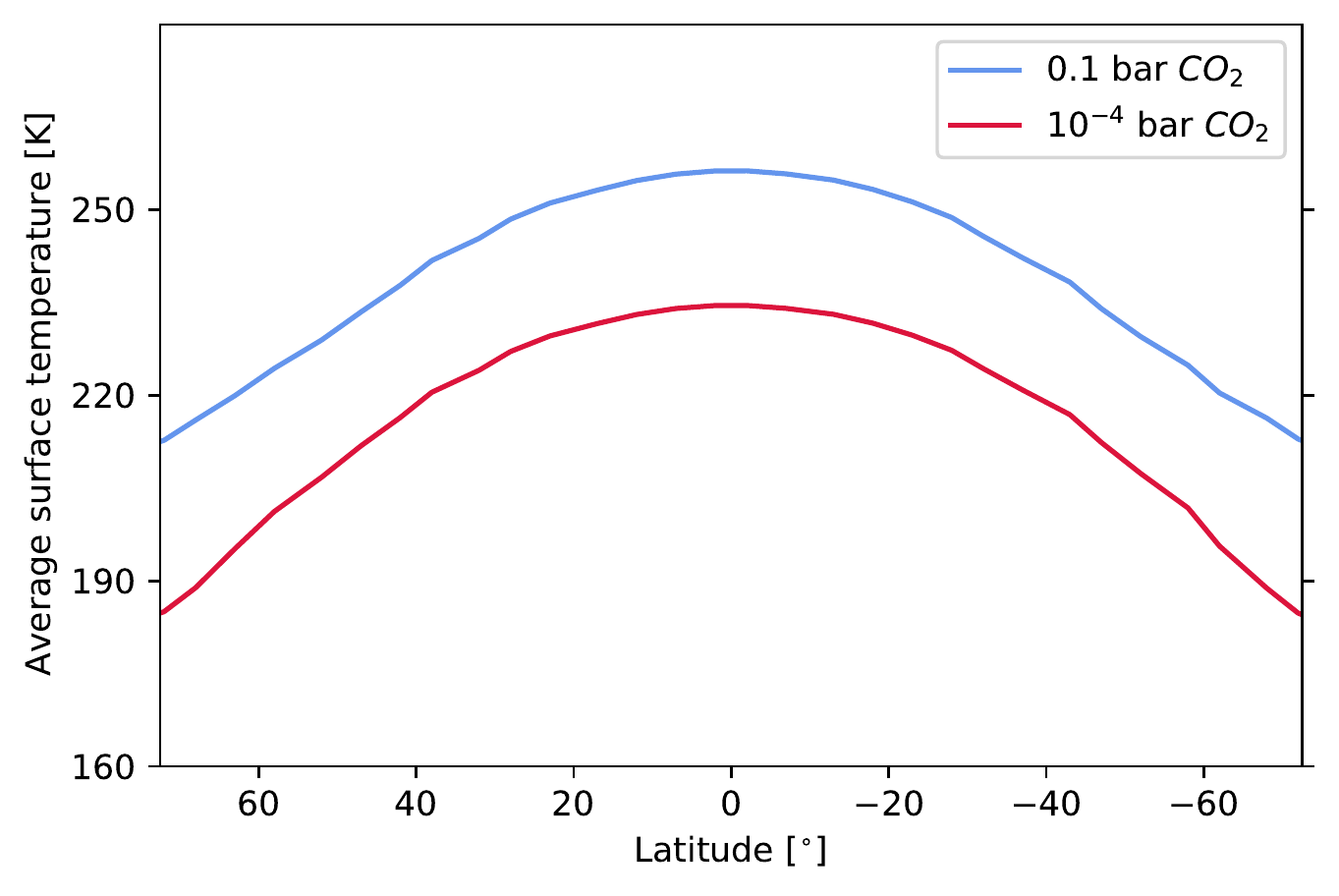}
    \caption{Average surface temperature as a function of latitude obtained using initial conditions corresponding to the GCM by \citet{hoffman_snowball_2017} (Figure 8a), for initial atmospheric $CO_{\rm 2}$ contents of $0.11$~mbar and $100$~mbar. The surface albedo was set to $0.6$ everywhere while planetary obliquity and eccentricity were set to $23.5^{\circ}$ and zero, respectively.}
    \label{fig:Hoffman_comparison}
\end{figure}

Surface condensation of $CO_{\mathrm{2}}$ ice is highly dependent on the equator-to-pole temperature contrast, which in turn is sensitive to the diffusion coefficient $D$ (Equation~\ref{hdiff}), which includes the planetary obliquity. We compared our results with those of past GCM simulations of snowball planets \citep{hoffman_snowball_2017} to test the reliability of our diffusion parameterization. Figure~\ref{fig:Hoffman_comparison} shows the predicted latitudinally-averaged surface temperature distributions in our model. In general, our results agree quite well with the ones of \citet{hoffman_snowball_2017}, lying near the lower range of surface temperatures, similar to the FOAM (Fast Ocean Atmosphere Model) model. Some of the GCM simulations are slightly warmer than ours (except for the FOAM model) because they predict the formation of highly-absorbing upper atmospheric $H_{\rm 2}O$ ice clouds in these cold and (otherwise) dry atmospheres. More models should test this result. Nevertheless, we conclude that our employed heat diffusion parameterization is reasonable and consistent with previous studies.

In summary, the main differences between EBMs and GCMs are related to how large scale dynamics (which might lead to sharper climatic transitions) and clouds are treated, whose spatial and vertical distributions are not computed in most EBMs. In spite of that, our EBM exhibits latitudinal mean surface temperatures for Earth that are consistent with real-world data \citep{RamirezLevi2018}. Our consistency with the GCM snowball simulations of \citet{hoffman_snowball_2017} is further suggestive of this. Furthermore, EBMs like the one presented here are computationally efficient, enabling the exploration of a wide parameter space, which would be computationally expensive with GCMs. Since this study mainly addresses different climatic trends and distributions, EBMs are appropriate tools. Ultimately, improved observations are needed to test how well climate models agree with reality, irrespective of complexity.

The habitable zone is the circumstellar region where standing bodies of liquid water could exist on the surface of a rocky planet \citep{Ramirez2018}. In our solar system, a conservative estimate spans from \textasciitilde $0.95$ - $1.67$ AU \citep{kasting1993, Ramirez2018}. Warm starts have traditionally been assumed in these
and most other calculated habitable zone limits \citep{kasting1993, pierregaidos2011, KumarKopparapu2013,zsom2013, ramirez2017, ramirez2020}. 
In contrast, \citet{kadoya_outer_2019} have shown that the habitable zone outer edge limit distance decreases for systems composed of cold start planets.  Although that conclusion is consistent with the trends we find here for cold and warm start planets across all distances (Figure~\ref{fig:ss_all}), we stress that a system's outer edge limit corresponds to that for warm start planets should at least one such planet exist. In other words, without additional information regarding the planets themselves, the absolute outer edge limit in a stellar system is best characterized by its warm start limit (as computed previously) \citep{kasting1993, KumarKopparapu2013}.

We have compared our cold and warm start habitable zone limits with those of \citet{kadoya_outer_2019}, assuming the stellar properties outlined in Table \ref{tab:parameters}. For the Sun, \citet{kadoya_outer_2019} obtain warm and cold start limits at $\sim1.69$ and $1.49$~AU, respectively, whereas corresponding limits for this study were found at $\sim 1.68$ and $1.39$~AU. We
attribute the slightly smaller cold start habitable zone limit ($\sim 7~\%$ decrease) to differences in ice-albedo feedback parameterizations and in our addition of clouds, which are not included in their model.
Likewise, their F-star warm and cold start limits are computed at $\sim 3.2$ and $2.75$~AU whereas our F0 limits are located at $\sim 3.1$ and $2.62$~AU, respectively. Therefore, the F0 values obtained here are smaller by about $3$-$5\%$. Again, our K5 warm and cold start limits are at $\sim 0.74$ and $0.65$~AU, respectively, as compared to their K-star values at 0.73 and 0.66 AU. These are within 2$\%$ of each other and are overall consistent.  We note that any remaining differences with \citet{kadoya_outer_2019} could be attributed to differences in the employed stellar spectra.

Similarly to \citet{kadoya_outer_2019}, we find that the difference in the stellar effective flux, $S_{\rm eff}$, between cold and warm start distances decreases from hotter to cooler stars. Assuming $S_{\rm eff}=1$ is the incident energy received on Earth at 1~AU from the Sun, $S_{\rm eff}$ can be computed following \citet{kasting1993} as

\begin{equation}
S_{\rm eff} = \frac{L}{d^{2}}.
\end{equation}

Here, $L$ is stellar luminosity in solar units and $d$ is the semi-major axis in astronomical units (AU). For instance, the solar $S_{\rm eff}$ from warm to cold starts increases by $\sim 0.17$ (i.e., from $\sim 0.35$ to $0.52$), whereas this increase was only $\sim 0.09$ (from 0.27 to 0.36) for K5 stars. The $S_{\rm eff}$ increase was largest at $\sim 0.18$ (0.447 to 0.63) for the hotter (F0) stars. This trend occurs because the ice-albedo feedback weakens, whereas near-infrared absorption intensifies for habitable zone planets orbiting cooler stars, reducing the difference in $S_{\rm eff}$ between cold and warm start limits. 

\section{Conclusions}
Planets similar to Earth might experience cold or warm stages during the course of their evolution. The carbonate-silicate cycle can stabilize planetary temperatures over geological timescales. While  $CO_{\mathrm{2}}$ generally increases the surface temperature, high enough atmospheric $CO_{\mathrm{2}}$ can trigger atmospheric collapse via $CO_{\mathrm{2}}$ surface ice condensation starting at the poles, potentially leading to irreversible glaciation. Such a process could negatively impact the habitability of terrestrial bodies, even if located within the canonical habitable zone. This was argued in \citet{Turbet2017}, who showed that surface $CO_{\mathrm{2}}$ ice condensation on initially frozen planets orbiting the Sun can significantly reduce the range of habitable orbital distances. 

Our work confirms these results and extends the analysis to rapidly-rotating planets under different starting temperature conditions orbiting a range of star types (F - K). Using a latitudinally-dependent EBM, we show that planets that start out warm generally exhibit surface $CO_{\mathrm{2}}$ ice condensation at significantly larger orbital distances than cold start (i.e., fully glaciated with $T_{\mathrm{surf}}=230$~K) planets. This implies a wide habitable zone, consistent with what had been previously computed using simpler 1-D models \citep{kasting1993,KumarKopparapu2013,Ramirez2018}. We also find that K-star planets are more resistant to atmospheric collapse than planets around larger stars. Finally, our heat diffusion parameterization provides a simple way of evaluating a range of warm and cold climates with minimal tuning of parameters. 

The physics of planetary atmospheres remains complex, particularly for planets unlike the Earth. Future work should continue to use a hierarchy of models to explore the effects of clouds, convection, and the impact that oceans have on the overall heat transport. 

\section*{Acknowledgements}
I.B. acknowledges financial support from the Japanese Ministry of Education, Culture, Sports, Science and Technology (MEXT) and the Japanese Society for the Promotion of Science (JSPS).
R.M.R. acknowledges funding from the Earth-Life Science Institute (ELSI) and from the National Institutes of Natural Sciences: Astrobiology Center (grant number JY310064). We thank the anonymous reviewer for their constructive comments, which considerably improved the manuscript.

\section*{Data availability}
The code and the jupyter notebooks used for data analysis are available on Github at https://github.com/irenebonati/CO2-condensation.

\bibliographystyle{mnras}
\bibliography{bibliography} % if your bibtex file is called example.bib

\begin{thebibliography}{}
\makeatletter
\relax
\def\mn@urlcharsother{\let\do\@makeother \do\$\do\&\do\#\do\^\do\_\do\%\do\~}
\def\mn@doi{\begingroup\mn@urlcharsother \@ifnextchar [ {\mn@doi@}
  {\mn@doi@[]}}
\def\mn@doi@[#1]#2{\def\@tempa{#1}\ifx\@tempa\@empty \href
  {http://dx.doi.org/#2} {doi:#2}\else \href {http://dx.doi.org/#2} {#1}\fi
  \endgroup}
\def\mn@eprint#1#2{\mn@eprint@#1:#2::\@nil}
\def\mn@eprint@arXiv#1{\href {http://arxiv.org/abs/#1} {{\tt arXiv:#1}}}
\def\mn@eprint@dblp#1{\href {http://dblp.uni-trier.de/rec/bibtex/#1.xml}
  {dblp:#1}}
\def\mn@eprint@#1:#2:#3:#4\@nil{\def\@tempa {#1}\def\@tempb {#2}\def\@tempc
  {#3}\ifx \@tempc \@empty \let \@tempc \@tempb \let \@tempb \@tempa \fi \ifx
  \@tempb \@empty \def\@tempb {arXiv}\fi \@ifundefined
  {mn@eprint@\@tempb}{\@tempb:\@tempc}{\expandafter \expandafter \csname
  mn@eprint@\@tempb\endcsname \expandafter{\@tempc}}}

\bibitem[\protect\citeauthoryear{Asolekar, Valentine  \& Schnoor}{Asolekar
  et~al.}{1991}]{asolekar1991}
Asolekar S.,  Valentine R.,   Schnoor J.~L.,  1991, Water resources research,
  27, 527

\bibitem[\protect\citeauthoryear{Batalha, Kopparapu, Haqq-Misra  \&
  Kasting}{Batalha et~al.}{2016}]{batalha2016}
Batalha N.~E.,  Kopparapu R.~K.,  Haqq-Misra J.,   Kasting J.~F.,  2016, Earth
  and Planetary Science Letters, 455, 7

\bibitem[\protect\citeauthoryear{Caballero \& Langen}{Caballero \&
  Langen}{2005}]{Caballero2005a}
Caballero R.,  Langen P.~L.,  2005, \mn@doi [Geophysical Research Letters]
  {10.1029/2004GL021581}, 32, 1

\bibitem[\protect\citeauthoryear{Forget, Hourdin  \& Talagrand}{Forget
  et~al.}{1998}]{forget1998}
Forget F.,  Hourdin F.,   Talagrand O.,  1998, Icarus, 131, 302

\bibitem[\protect\citeauthoryear{Forget, Wordsworth, Millour, Madeleine,
  Kerber, Leconte, Marcq  \& Haberle}{Forget et~al.}{2013}]{forget2013}
Forget F.,  Wordsworth R.,  Millour E.,  Madeleine J.-B.,  Kerber L.,  Leconte
  J.,  Marcq E.,   Haberle R.~M.,  2013, Icarus, 222, 81

\bibitem[\protect\citeauthoryear{Fray \& Schmitt}{Fray \&
  Schmitt}{2009}]{fray_sublimation_2009}
Fray N.,  Schmitt B.,  2009, \mn@doi [Planetary and Space Science]
  {10.1016/j.pss.2009.09.011}, 57, 2053

\bibitem[\protect\citeauthoryear{Graham \& Pierrehumbert}{Graham \&
  Pierrehumbert}{2020}]{graham-a}
Graham R.,  Pierrehumbert R.,  2020, Astrophysical Journal, 896

\bibitem[\protect\citeauthoryear{Haqq-Misra, Kopparapu, Batalha, Harman  \&
  Kasting}{Haqq-Misra et~al.}{2016}]{haqq2016limit}
Haqq-Misra J.,  Kopparapu R.~K.,  Batalha N.~E.,  Harman C.~E.,   Kasting
  J.~F.,  2016, The Astrophysical Journal, 827, 120

\bibitem[\protect\citeauthoryear{Hayworth, Kopparapu, Haqq-Misra, Batalha,
  Payne, Foley, Ikwut-Ukwa  \& Kasting}{Hayworth
  et~al.}{2020}]{hayworth_warming_2020}
Hayworth B. P.~C.,  Kopparapu R.~K.,  Haqq-Misra J.,  Batalha N.~E.,  Payne
  R.~C.,  Foley B.~J.,  Ikwut-Ukwa M.,   Kasting J.~F.,  2020, \mn@doi [Icarus]
  {10.1016/j.icarus.2020.113770}, 345, 113770

\bibitem[\protect\citeauthoryear{Hoffman, Kaufman, Halverson  \&
  Schrag}{Hoffman et~al.}{1998}]{Hoffman1342}
Hoffman P.~F.,  Kaufman A.~J.,  Halverson G.~P.,   Schrag D.~P.,  1998, \mn@doi
  [Science] {10.1126/science.281.5381.1342}, 281, 1342

\bibitem[\protect\citeauthoryear{Hoffman et~al.,}{Hoffman
  et~al.}{2017}]{hoffman_snowball_2017}
Hoffman P.~F.,  et~al., 2017, \mn@doi [Science Advances]
  {10.1126/sciadv.1600983}, 3, e1600983

\bibitem[\protect\citeauthoryear{James \& North}{James \&
  North}{1982}]{james1982}
James P.~B.,  North G.~R.,  1982, Journal of Geophysical Research: Solid Earth,
  87, 10271

\bibitem[\protect\citeauthoryear{Kadoya \& Tajika}{Kadoya \&
  Tajika}{2019}]{kadoya_outer_2019}
Kadoya S.,  Tajika E.,  2019, \mn@doi [The Astrophysical Journal]
  {10.3847/1538-4357/ab0aef}, 875, 7

\bibitem[\protect\citeauthoryear{Kasting}{Kasting}{1991}]{Kasting1991}
Kasting J.~F.,  1991, \mn@doi [Icarus] {10.1016/0019-1035(91)90137-I}, 94, 1

\bibitem[\protect\citeauthoryear{Kasting}{Kasting}{2014}]{kasting2014}
Kasting J.~F.,  2014, Geological Society of America Special Papers, 504, 19

\bibitem[\protect\citeauthoryear{Kasting, Whitmire  \& Reynolds}{Kasting
  et~al.}{1993}]{kasting1993}
Kasting J.~F.,  Whitmire D.~P.,   Reynolds R.~T.,  1993, Icarus, 101, 108

\bibitem[\protect\citeauthoryear{Kirschvink}{Kirschvink}{1992}]{kirschvink1992}
Kirschvink J.~L.,  1992, Late Proterozoic low-latitude global glaciation: the
  snowball Earth.
Cambridge University Press

\bibitem[\protect\citeauthoryear{Kite, Mischna, Gao, Yung  \& Turbet}{Kite
  et~al.}{2020}]{kite_methane_2020}
Kite E.~S.,  Mischna M.~A.,  Gao P.,  Yung Y.~L.,   Turbet M.,  2020, \mn@doi
  [P\&SS] {10.1016/j.pss.2019.104820}, 181, 104820

\bibitem[\protect\citeauthoryear{Kitzmann}{Kitzmann}{2016}]{kitzmann2016}
Kitzmann D.,  2016, The Astrophysical Journal Letters, 817, L18

\bibitem[\protect\citeauthoryear{{Kopparapu} et~al.,}{{Kopparapu}
  et~al.}{2013}]{KumarKopparapu2013}
{Kopparapu} R.~K.,  et~al., 2013, \mn@doi [The Astrophysical Journal]
  {10.1088/0004-637X/765/2/131}, 765, 131

\bibitem[\protect\citeauthoryear{North \& Coakley~Jr}{North \&
  Coakley~Jr}{1979}]{North1979}
North G.~R.,  Coakley~Jr J.~A.,  1979, Journal of the Atmospheric Sciences, 36,
  1189

\bibitem[\protect\citeauthoryear{North, Cahalan  \& Coakley}{North
  et~al.}{1981}]{North1981}
North G.~R.,  Cahalan R.~F.,   Coakley J.~A.,  1981, Nasa-Cr-177133

\bibitem[\protect\citeauthoryear{Paradise \& Menou}{Paradise \&
  Menou}{2017}]{paradise2017}
Paradise A.,  Menou K.,  2017, The Astrophysical Journal, 848, 33

\bibitem[\protect\citeauthoryear{Pierrehumbert}{Pierrehumbert}{2005}]{Phumbert2005}
Pierrehumbert R.~T.,  2005, \mn@doi [Journal of Geophysical Research:
  Atmospheres] {10.1029/2004JD005162}, 110

\bibitem[\protect\citeauthoryear{Pierrehumbert \& Gaidos}{Pierrehumbert \&
  Gaidos}{2011}]{pierregaidos2011}
Pierrehumbert R.,  Gaidos E.,  2011, The Astrophysical Journal Letters, 734,
  L13

\bibitem[\protect\citeauthoryear{Pierrehumbert, Abbot, Voigt  \&
  Koll}{Pierrehumbert et~al.}{2011}]{Phumbert2011}
Pierrehumbert R.,  Abbot D.,  Voigt A.,   Koll D.,  2011, \mn@doi [Annual
  Review of Earth and Planetary Sciences]
  {10.1146/annurev-earth-040809-152447}, 39, 417

\bibitem[\protect\citeauthoryear{Ramirez}{Ramirez}{2017}]{ramirez2017mars}
Ramirez R.~M.,  2017, Icarus, 297, 71

\bibitem[\protect\citeauthoryear{Ramirez}{Ramirez}{2018}]{Ramirez2018}
Ramirez R.~M.,  2018, Geosciences, 8, 280

\bibitem[\protect\citeauthoryear{Ramirez}{Ramirez}{2020}]{ramirez2020}
Ramirez R.~M.,  2020, Monthly Notices of the Royal Astronomical Society, 494,
  259

\bibitem[\protect\citeauthoryear{Ramirez \& Kaltenegger}{Ramirez \&
  Kaltenegger}{2017}]{ramirez2017}
Ramirez R.~M.,  Kaltenegger L.,  2017, The Astrophysical Journal Letters, 837,
  L4

\bibitem[\protect\citeauthoryear{Ramirez \& Kaltenegger}{Ramirez \&
  Kaltenegger}{2018}]{ramirezkalt2018}
Ramirez R.~M.,  Kaltenegger L.,  2018, The Astrophysical Journal, 858, 72

\bibitem[\protect\citeauthoryear{Ramirez \& Levi}{Ramirez \&
  Levi}{2018}]{RamirezLevi2018}
Ramirez R.~M.,  Levi A.,  2018, \mn@doi [Monthly Notices of the Royal
  Astronomical Society] {10.1093/mnras/sty761}, 4640, 4627

\bibitem[\protect\citeauthoryear{Ramirez, Kopparapu, Zugger, Robinson, Freedman
   \& Kasting}{Ramirez et~al.}{2014}]{ramirez2014}
Ramirez R.~M.,  Kopparapu R.,  Zugger M.~E.,  Robinson T.~D.,  Freedman R.,
  Kasting J.~F.,  2014, Nature Geoscience, 7, 59

\bibitem[\protect\citeauthoryear{Ramirez, Craddock  \& Usui}{Ramirez
  et~al.}{2020}]{ramirezetal2020}
Ramirez R.~M.,  Craddock R.~A.,   Usui T.,  2020, Journal of Geophysical
  Research: Planets, 125, e2019JE006160

\bibitem[\protect\citeauthoryear{Rose, Cronin  \& Bitz}{Rose
  et~al.}{2017}]{Rose2017}
Rose B. E.~J.,  Cronin T.~W.,   Bitz C.~M.,  2017, \mn@doi [The Astrophysical
  Journal] {10.3847/1538-4357/aa8306}, 846, 28

\bibitem[\protect\citeauthoryear{Soto, Mischna, Schneider, Lee  \&
  Richardson}{Soto et~al.}{2015}]{Soto2015}
Soto A.,  Mischna M.,  Schneider T.,  Lee C.,   Richardson M.,  2015, \mn@doi
  [Icarus] {10.1016/j.icarus.2014.11.028}, 250, 553

\bibitem[\protect\citeauthoryear{{Turbet}, {Forget}, {Leconte}, {Charnay}  \&
  {Tobie}}{{Turbet} et~al.}{2017}]{Turbet2017}
{Turbet} M.,  {Forget} F.,  {Leconte} J.,  {Charnay} B.,   {Tobie} G.,  2017,
  \mn@doi [Earth and Planetary Science Letters] {10.1016/j.epsl.2017.07.050},
  \href {https://ui.adsabs.harvard.edu/abs/2017E&PSL.476...11T} {476, 11}

\bibitem[\protect\citeauthoryear{Turbet, Tran, Pirali, Forget, Boulet  \&
  Hartmann}{Turbet et~al.}{2019}]{turbet_far_2019}
Turbet M.,  Tran H.,  Pirali O.,  Forget F.,  Boulet C.,   Hartmann J.-M.,
  2019, \mn@doi [Icarus] {10.1016/j.icarus.2018.11.021}, 321, 189

\bibitem[\protect\citeauthoryear{Villanueva et~al.,}{Villanueva
  et~al.}{2015}]{villanueva2015}
Villanueva G.,  et~al., 2015, Science, 348, 218

\bibitem[\protect\citeauthoryear{Vladilo, Murante, Silva, Provenzale, Ferri  \&
  Ragazzini}{Vladilo et~al.}{2013}]{vladilo2013}
Vladilo G.,  Murante G.,  Silva L.,  Provenzale A.,  Ferri G.,   Ragazzini G.,
  2013, The Astrophysical Journal, 767, 65

\bibitem[\protect\citeauthoryear{Warren, Wiscombe  \& Firestone}{Warren
  et~al.}{1990}]{warren_spectral_1990}
Warren S.~G.,  Wiscombe W.~J.,   Firestone J.~F.,  1990, \mn@doi [Journal of
  Geophysical Research: Solid Earth] {10.1029/JB095iB09p14717}, 95, 14717

\bibitem[\protect\citeauthoryear{Williams \& Kasting}{Williams \&
  Kasting}{1997}]{Williams1997}
Williams D.~M.,  Kasting J.~F.,  1997, \mn@doi [Icarus]
  {10.1006/icar.1997.5759}, 129, 254

\bibitem[\protect\citeauthoryear{Williams \& Pollard}{Williams \&
  Pollard}{2003}]{WilliamsPollard}
Williams D.,  Pollard D.,  2003, \mn@doi [International Journal of Astrobiology
  - INT J ASTROBIOL] {10.1017/S1473550403001356}, 2

\bibitem[\protect\citeauthoryear{Wolf}{Wolf}{2018}]{wolf2018erratum}
Wolf E.~T.,  2018, Astrophys. J. Lett, 855

\bibitem[\protect\citeauthoryear{Wordsworth, Forget, Selsis, Madeleine, Millour
   \& Eymet}{Wordsworth et~al.}{2010}]{wordsworth2010}
Wordsworth R.,  Forget F.,  Selsis F.,  Madeleine J.-B.,  Millour E.,   Eymet
  V.,  2010, Astronomy \& Astrophysics, 522, A22

\bibitem[\protect\citeauthoryear{Wordsworth, Forget, Millour, Head, Madeleine
  \& Charnay}{Wordsworth et~al.}{2013}]{wordsworth2013}
Wordsworth R.,  Forget F.,  Millour E.,  Head J.,  Madeleine J.-B.,   Charnay
  B.,  2013, Icarus, 222, 1

\bibitem[\protect\citeauthoryear{Wordsworth, Kalugina, Lokshtanov, Vigasin,
  Ehlmann, Head, Sanders  \& Wang}{Wordsworth
  et~al.}{2017}]{wordsworth_transient_2017}
Wordsworth R.,  Kalugina Y.,  Lokshtanov S.,  Vigasin A.,  Ehlmann B.,  Head
  J.,  Sanders C.,   Wang H.,  2017, \mn@doi [Geophysical Research Letters]
  {10.1002/2016GL071766}, 44, 665

\bibitem[\protect\citeauthoryear{Xu \& Krueger}{Xu \& Krueger}{1991}]{xu1991}
Xu K.-M.,  Krueger S.~K.,  1991, Monthly weather review, 119, 342

\bibitem[\protect\citeauthoryear{Yang \& Abbot}{Yang \& Abbot}{2014}]{yang2014}
Yang J.,  Abbot D.~S.,  2014, The Astrophysical Journal, 784, 155

\bibitem[\protect\citeauthoryear{Zsom, Seager, De~Wit  \& Stamenkovi{\'c}}{Zsom
  et~al.}{2013}]{zsom2013}
Zsom A.,  Seager S.,  De~Wit J.,   Stamenkovi{\'c} V.,  2013, The Astrophysical
  Journal, 778, 109

\makeatother
\end{thebibliography}

% Don't change these lines
\bsp	% typesetting comment
\label{lastpage}
\end{document}